\documentclass[10pt,letterpaper,reprint,graphicx,superscriptaddress]{article}
\usepackage{opex3}
\usepackage{cite}

\begin{document}

\title{Intense isolated few-cycle attosecond XUV pulses from overdense plasmas driven by tailored laser pulses}

\author{Zi-Yu Chen,$^{1,2,*}$ Xiao-Ya Li,$^{1}$ Li-Ming Chen,$^{2}$ Yu-Tong Li,$^{2}$ and Wen-Jun Zhu$^{1}$}

\address{$^{1}$National Key Laboratory of Shock Wave and Detonation Physics, Institute of Fluid Physics, China Academy of Engineering Physics,
Mianyang, Sichuan 621900, China \\ $^{2}$Beijing National Laboratory of Condensed Matter Physics, Institute of Physics, Chinese Academy of Sciences,
Beijing 100190, China}

\email{$^{*}$ziyuch@caep.ac.cn} 



\begin{abstract}
A new scheme to generate an intense isolated few-cycle attosecond XUV pulse is demonstrated using particle-in-cell simulations. By use of unipolarlike or subcycle laser pulses irradiating a thin foil target, a strong transverse net current can be excited, which emits a few-cycle XUV pulse from the target rear side. The isolated pulse is ultrashort in the time domain with duration of several hundred attoseconds. The pulse also has a narrow bandwidth in the spectral domain compared to other XUV sources of high-order harmonics. It has most energy confined around the plasma frequency and no low-harmonic orders below the plasma frequency. It is also shown that XUV pulse of peak field strength up to $ 8\times 10^{12} $ V$\mathrm{m}^{-1}$ can be produced. Without the need for pulse selecting and spectral filtering, such an intense few-cycle XUV pulse is better suited to a number of applications.
\end{abstract}



\section{Introduction}

Laser-driven coherent XUV radiation has become a dynamic research topic in the past decade \cite{Winterfeldt2008,Teubner2009}. This new type of light source is promising for numerous attractive applications. Firstly, these laser-driven XUV pulses usually emit in the form of high-order harmonics with broadband spectrum, corresponding to ultrashort (femto-attosecond region) pulse durations in the time domain. This opens a door to investigate ultrafast dynamics and explore novel attosecond physics \cite{Krausz2009}. As the workhorses of current attosecond science, the generation of attosecond XUV pulses enables real-time tracking of the ultrafast process of electron dynamics outside the atomic nucleus. Besides, intense coherent XUV pulses offer a new opportunity on nonlinear optics in the XUV region. Various new phenomena may be found in this region that are quite different from those in optical field, such as strong field ionization in higher frequencies and ultrafast spectroscopy at shorter wavelengths \cite{Zhang2010,Heissler2012a}. In addition, XUV pulses can also be used as unique diagnostics which shed light on laser-matter interaction physics \cite{Borot2012}, including the underlying physics of radiation process itself and the properties of dense plasmas \cite{Dobosz2005}.

High-order harmonics generation (HHG) in atomic gases has been explored extensively as a route to produce attosecond XUV pulses \cite{Winterfeldt2008}. The emission process can be well understood in the framework of the three step model \cite{Corkum1993}. The main concern for such XUV pulses is the typically rather weak strength and low generation efficiency, due to the limitation of medium depletion through ionization. Many applications require substantially increasing the XUV brightness. To achieve high intensity, laser-plasma interaction has been exploited as a promising alternative because plasma can withstand high laser fields driving the harmonics \cite{Teubner2009}. Two distinct generation mechanisms have been identified to contribute to HHG from solid plasmas: coherent wake emission (CWE) \cite{Quere2006,Thaury2007,Nomura2008,Wheeler2012} and the relativistic oscillating mirror (ROM) process \cite{Bulanov1994,Lichters1996,Baeva2006,Dromey2006}. Both mechanisms emit XUV radiation in the reflected direction through nonlinear conversion processes at the plasma front surfaces. Another mechanism named coherent synchrotron emission (CSE) by dense electron nanobunches can generate coherent XUV pulses in the transmitted direction \cite{Brugge2010,Dromey2012}. These types of XUV pulses, however, are usually structured as a comb of harmonics with broad bandwidth in the spectral domain and as a train of attosecond pulse bursts in the time domain. In order to get ultrashort attosecond pulses, low-harmonic orders have to be filtered out. To obtain single isolated attosecond pulse, techniques such as polarization gating \cite{Baeva2006,Rykovanov2008} and few-cycle driving laser pulses \cite{Heissler2012,Liu2008} have been investigated, because isolated attosecond pulse is highly desired for many special applications, for example, time-resolved imaging and spectroscopy.

In this paper, we propose a very different route to produce intense attosecond XUV pulses from overdense plasmas. By using tailored laser pulse targeting onto a 100 nm thick plasma layer, an isolated few-cycle attosecond XUV pulse can be emitted from the target rear side. The pulse has attractive characteristics both in the temporal and spectral domain, which is better suited to a number of applications such as ultrafast dynamics, diagnostic of dense plasma properties, and nonlinear optics in the XUV region.

\section{Scheme and parameters}

\begin{figure}
\centering\includegraphics[width=0.7\textwidth
]{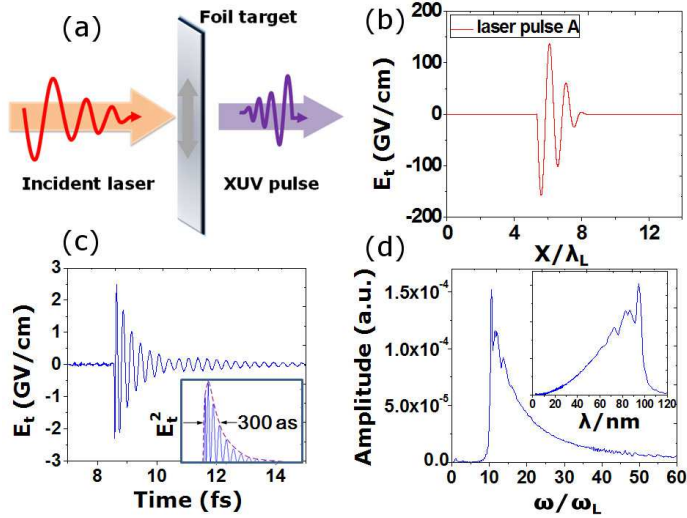}
\caption{\label{scheme} (a) Schematic diagram of the interaction. (b) The initial temporal profile of a sharply decreasing unipolarlike laser pulse, referred to as laser pulse A. (c) The temporal evolution profile of the XUV pulse observed at the rear side of the plasma irradiated by laser pulse A. The inset shows the temporal profile plotted in terms of $ E^2_t $. (d) The corresponding Fourier spectra of the XUV pulse. The inset shows the spectra plotted in terms of wavelength.}
\end{figure}

We performed one-dimensional (1D) particle-in-cell (PIC) simulations with the code VORPAL \cite{Nieter2004} to identify the XUV pulse generation mechanism. The PIC method is a basic plasma simulation technique which simulates a plasma system by following the trajectories of a number of charged particles. The method is based on self-consistently solving Maxwell’s equations determining the electric and magnetic fields and the equation of motion for the macroparticles. It should be noted that 1D model is more applicable for larger laser spot size as it does not taken multidimensional effects into account, such as hole boring and Rayleigh-Taylor-like instabilities. The schematic of the interaction is shown in Fig. \ref{scheme}(a). A plane femtosencond laser pulse propagates along the $ x $-direction and normally incidents onto a solid foil target. The laser pulse is linearly polarized along the $ z $-direction with a sine square amplitude profile $a_z=eE_z/m_e\omega_Lc=a_L\sin^2(\pi t/\tau_L)\sin(\omega_L t)\hat{z}$, where $ E_z $ is the laser electric field, $ \omega_L $ is the laser angular frequency, $\tau_L$ is the full laser pulse duration, $a_L$ is the peak amplitude of the normalized vector potential $ a_z $, $e$ is the electron charge, $ m_e $ is the electron mass, and $ c $ is the light speed in vacuum. The laser field strength $a_L=5$ corresponds to an intensity of $ I=3.4\times 10^{19} \mathrm{W} /\mathrm{cm}^2$ through the relation $I\lambda_L^2=a_L^2\times1.37\times10^{18} \mathrm{W} \mu \mathrm{m}^2/\mathrm{cm}^2$ for a laser of wavelength $ \lambda_L=1\mu \mathrm{m} $.

The key element of this scheme is to use laser pulse with a sharp edge \cite{Wang2011,Li2013,Ji2009,Chen2013}. Both sharply increasing and decreasing laser pulses give rise to the same emission. To illustrate the emission characteristic more clearly, we firstly consider laser pulses with an extremely sharp decreasing edge, i.e, unipolarlike pulses \cite{Rau1997}. Figure \ref{scheme}(b) shows the profile of such a laser pulse with $\tau_L=3T_L$ (referred to as laser pulse A), where $ T_L $ is the laser oscillating period.

The foil target is taken to be fully ionized plasma with a uniform density of $ n_e=100 n_c $, where $ n_c= \omega_L^2 m_e/4\pi e^2 $ is the critical density for the laser. Such an electron density corresponding to an electron plasma frequency of $\omega_p = \sqrt{4\pi n_e e^2/m_e}=10 \omega_L$. The foil thickness is chosen to be $ d=150 $nm, which is comparable to the plasma wavelength $ \lambda_p=100 $nm. Therefore radiation with frequencies above $ \omega_p $ can tunnel through the thin plasma and radiate into vacuum \cite{Wu2008}, while the laser pulses and frequencies blew $ \omega_p $ can not penetrate through the foil. Such thin foils can be made of diamond-like carbon foils. For the ultrathin target foils, high-contrast laser pulses are required to ensure the target not destroyed or deformed at the arrival of the main pulse. Such clean pulses are presently available in experiments \cite{Kiefer2013}. The ions are set immobile because of the short time of interaction. To resolve the attosecond pulse properly,  the cell size is $ \lambda/1000 $ and each cell is filled with 100 macroparticles.

\section{Results and discussions}

Figure \ref{scheme}(c) shows the temporal profile of the XUV pulse observed from the rear side of the foil target. The XUV emission is polarized in the $ z $-direction, the same as the laser pulse. The XUV pulse lasts only several periods and decays with time. The peak strength of the electric field $ E_t $ is found to be about 2.5 GV/cm, which is about two orders of magnitude smaller than the laser electric field strength of 150 GV/cm. The XUV pulse has a very sharp rising edge, which is demanding for experiments. The inset of Fig. \ref{scheme}(c) shows the temporal profile of $ E_t^2 $. The XUV pulse duration (full width at half maximum, FWHM) is about 300 as when plotted as intensity. In addition, the XUV pulse is an isolated attosecond pulse, compared to a train of attosecond pulses formed in other XUV sources. This feature is important for many experimental applications.

\begin{figure}
\centering\includegraphics[width=0.7\textwidth
]{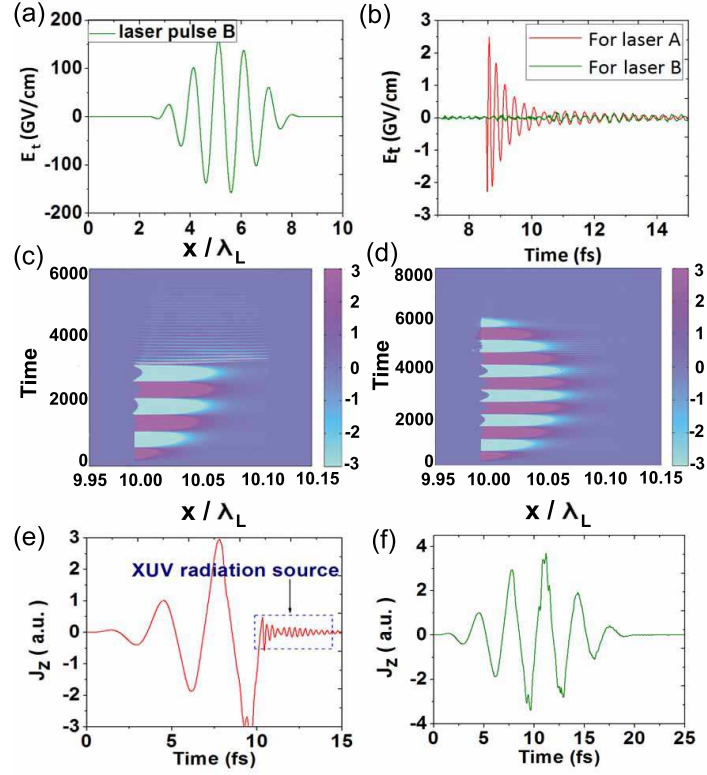}
\caption{\label{Jz} (a) The initial temporal profile of a sine squared laser pulse, referred to as laser pulse B. (b) The temporal profiles of the XUV pulses driven by laser pulse A (red) and laser pulse B (green), respectively. (c)-(d) The spatial-temporal distribution of the transverse currents $ J_z $ excited by laser pulse A (c) and laser pulse B (d), respectively. The foil target is initially located in the region between $ x=10.00\lambda_L $ and $ x=10.15\lambda_L $. (e)-(f) $ J_z $ plotted versus time along central line $ x=10.05 \lambda_L $ corresponding to (c) and (d), respectively.}
\end{figure}

The Fourier spectra of the XUV pulse is shown in Fig. \ref{scheme}(d). The central frequency of the XUV spectra is about $ \omega=10\omega_L $, which is just the initial plasma frequency $ \omega_p $. The spectra also contains high frequencies up to several tens of $ \omega_p $. This might be caused by the increase of local plasma frequency $ \omega_p(x) $, due to compression of the electron surface layer by the laser light pressure, as will be shown later. Frequencies lower than $ \omega_p $ vanish sharply in the spectra. As a result, we are able to produce a single attosecond pulse having clean frequency spectra in the XUV range without spectra filtering. This striking spectral feature can be distinguished from other different HHG generation mechanisms, i.e., CWE, ROM and CSE process. In the cases of CWE and ROM process, the emitted spectrum carries a portion of energy within the lower order harmonics up to the plasma frequency $ \omega_p $. The XUV spectra plotted versus wavelength is shown in the inset of Fig. \ref{scheme}(d). As can be seen, XUV pulse with wavelength in the 20-100 nm range has been obtained. The spectral bandwidth (FWHM) is about 30 nm. While processes of ROM and CSE both generate radiation with broadband spectrum extending to frequencies much higher than $ \omega_p $.

For the case of a laser pulse with symmetric sine square pulse shape, we plot the laser amplitude profile with $\tau_L=6T_L$ (referred to as laser pulse B) in Fig. \ref{Jz}(a). The corresponding temporal profile of XUV emission for laser pulse B is shown in Fig. \ref{Jz}(b). The XUV profile for laser pulse A is shown for comparison. Compared with the laser pulse with a sharp edge, the symmetric sine square laser pulse can only produce very weak XUV emission from the rear side of the foil. It is worth noting that we have also done simulations with laser pulses with a sharply rising edge. The generated XUV pulse is similar to that of the sharply decreasing laser pulses. This characteristic of emission also presents a route to identify unipolar half-cycle pulse \cite{Wu2012}.

\begin{figure}
\centering\includegraphics[width=0.7\textwidth
]{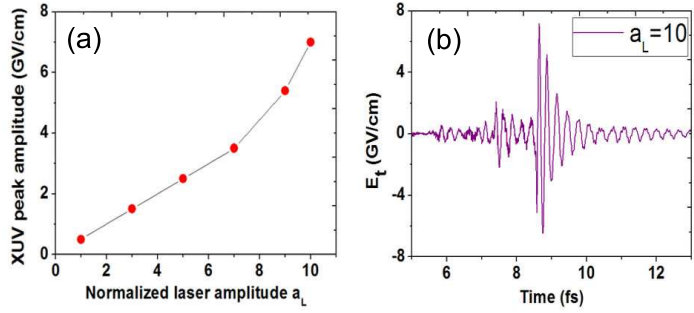}
\caption{\label{changea} (a) Peak XUV field strength as a function of laser amplitude from PIC simulations. (b) The temporal profile of XUV pulse using laser pulse with amplitude of $ a_L=10 $. The other parameters are the same as in Fig. \ref{scheme}.}
\end{figure}

To illustrate the underlying radiation mechanism, we display the temporal and spacial distribution of the transverse currents $ J_z $ for the two laser shapes in Figs. \ref{Jz}(c) and \ref{Jz}(d). The foil target is initially located in the region between $ x=10.00\lambda_L $ and $ x=10.15\lambda_L $. As can be seen, the electron layer on the target front surface has been compressed and pushed inward by the laser light pressure during the interaction process. For laser pulse A with a sharp edge, a strong transverse transient net current is formed after the laser pulse. The net current has a high-frequency wakefield oscillating profile modulated at the plasma frequency and decays with time (see Fig. \ref{Jz}(c)). While for symmetric laser pulse B, only weak transverse net current is formed after the laser pulse (see Fig. \ref{Jz}(d)). To see this more clearly, we have plotted $ J_z $ versus time along central line $ x=10.05 \lambda_L $ corresponding to Figs. \ref{Jz}(c) and \ref{Jz}(d), respectively (see Figs. \ref{Jz}(e) and \ref{Jz}(f)). The transverse transient net current is the radiation source responsible for the $ z $-polarized XUV emission at the plasma frequency.

To get a physical understanding of the origin of the transverse net current, we can employ a simple picture to show how the current can be excited by such laser pulses. The electrons do quiver motion under the action of the laser field analogy to forced oscillators. The transverse momentum can be obtained by integrating the $z$ component of the relativistic Lorentz equation to be \cite{Rau1997,Khachatryan2004}
\begin{equation}
p_z = A = e/(m_0 c^2) \int_{-\infty}^{\infty} E_z(\xi)d \xi,
\end{equation}
where $A$ is the vector potential and $ \xi =x-ct $. For laser pulses with a sharp edge such that high intensities are reached or decreased within less than two cycles, $A$ does not vanish \cite{Rau1997,Kundu2012}. That means, for unipolarlike or subcyclic pulse, the Lawson-Woodward theorem does not apply \cite{Rau1997}. And hence net energy gain happens and electron acceleration takes place. Therefore, the electrons will receive a kick in the transverse polarization direction after the laser pulse is over. Thereafter the electrons will keep on free oscillation with its natural frequency $ \omega_p $. As a result, a transverse net current is formed which emits electromagnetic pulse. Because the plasma width is limited which is comparable to the XUV wavelength, coherent radiation also at the plasma frequency will be emitted through the thin plasma. Lower-harmonic orders generated at the target front surface can not penetrate through the overcritical plasma. The few-cycle shape of the emitted pulse is mainly due to radiation damping effect.

The XUV field is proportional to the derivation of current density as $ E_{XUV} \propto \partial J_{z} / \partial t $. When the laser intensity is not too high to change the plasma density significantly, we can approximately have $ \partial J_{z} / \partial t \propto \partial p_{z}/\partial t $. Thus we can obtain the maximum XUV field scaling as $ E_{XUV} \propto a_{L} $. We have done a serial of simulations to study the XUV field strength as a function of the laser amplitude, as shown in Fig. \ref{changea}(a). The other parameters are the same as in Fig. \ref{scheme}. As is seen, when the laser intensity is relatively small, the XUV peak field amplitude does increase linearly with increasing the laser amplitude. While for greater laser amplitude, the slope shows a nonlinear trend. At the same time, however, the ultrashort characteristic of the XUV pulse becomes worse in quality (see Fig. \ref{changea}(b)). This is due to plasma oscillation becoming violent and irregular, which is not suitable to produce the shaped attosecond pulse. In the linear region, the energy conversion efficiency of laser to XUV pulse can be obtained to be more than $1\times10^{-5}$.

\begin{figure}
\centering\includegraphics[width=0.8\textwidth
]{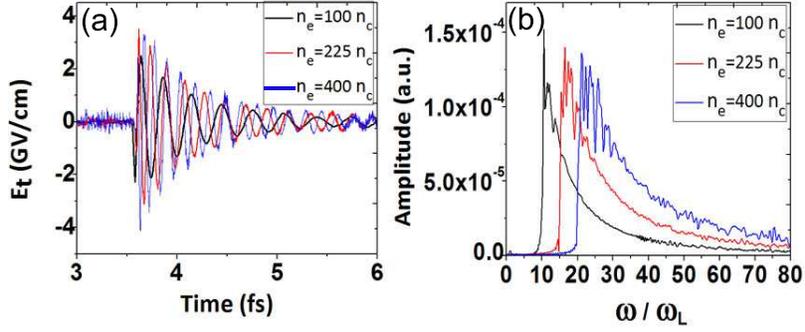}
\caption{\label{ne} (a) The temporal profiles and (b) the corresponding Fourier spectrum of the XUV pulses generated from three plasma slabs with different initial plasma densities: $ n_e =100 n_c$ (black), $ n_e =225 n_c$ (red), and $ n_e =400 n_c$ (blue). At the same time, the laser intensities are chosen to keep the dimensionless similarity parameter $ S=n_e/a_L n_c $ unchanged.}
\end{figure}

Since the XUV pulse emits mainly at the plasma frequency which is dependent on the plasma density, it provides a way to tune the XUV frequency spectra by changing the plasma density. Figures \ref{ne}(a) and \ref{ne}(b) display the temporal profiles and the corresponding Fourier spectrum of the XUV pulses for three different plasma densities $n_e=100n_c, 225n_c$, and $400n_c$. The corresponding plasma frequency is respectively $\omega_p=10\omega_L, 15\omega_L$, and $20\omega_L$. The laser amplitude is chosen to be $ a_L=5, 11.25 $, and $ 20 $, respectively. According to the similarity theory in the ultrarelativistic regime ($ a_L^2\gg 1 $), we keep the dimensionless similarity parameter $ S=n_e/a_L n_c $ the same so that the laser plasma dynamics remains similar \cite{Gordienko2005}. From Fig. \ref{ne}(a) we can see that the field strength and pulse duration of the three XUV pules are similar. For the Fourier spectrum shown in Fig. \ref{ne}(b), it is easy to find that the lower cut off frequencies are just at the plasma frequencies for different densities as expected. This feature of radiation also offers a new window on diagnosing the density of dense plasmas.

\begin{figure}
\centering\includegraphics[width=0.7\textwidth
]{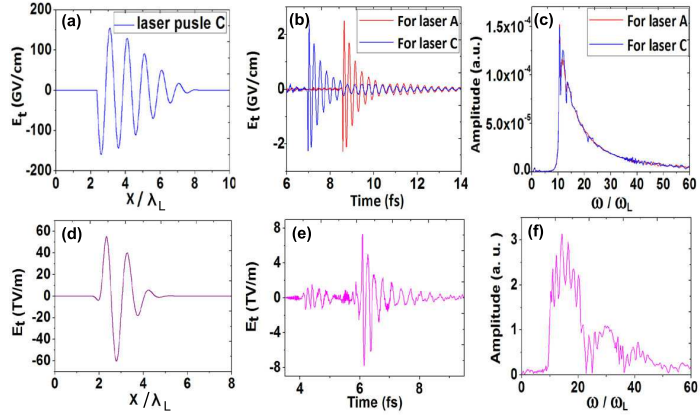}
\caption{\label{tau} (a) The initial temporal profile of a sharply decreasing unipolarlike laser pulse with full duration of $ \tau_L=6T_L $, referred to as laser pulse C. (b) The temporal profiles and (c) the corresponding Fourier spectrum of the XUV pulses generated by laser pulse A (red) and laser pulse C (blue). (d) The initial temporal profile of a quasi-single-cycle laser pulse based on Ref. \cite{Ji2009}, referred to as laser pulse D. (e) The temporal profiles and (f) the Fourier spectra of the XUV pulse generated by laser pulse D.}
\end{figure}

It should be pointed out that the present mechanism should work well for laser pulses with longer durations, as long as a sharp edge is kept such that high intensities are reached or decreased within less than two cycles. As a reference case to demonstrate this, we keep the laser pulse shape and peak amplitude the same as laser pulse A, while increasing the laser pulse duration to $\tau_L=6T_L$. This unipolarlike laser pulse is referred to as laser pulse C (see Fig. \ref{tau}(a)). As is seen in Figs. \ref{tau}(b) and \ref{tau}(c), both the temporal profile and the Fourier spectra of the XUV pulse are not affected by only changing the laser pulse duration.

Besides the unipolarlike laser pulses, ultrashort laser pulse with less than two cycles can also be used. Here we consider a quasi-single-cycle relativistic laser pulse with profile similar to that generated by the method proposed by Ji \textit{et al}. in Ref. \cite{Ji2009} (see Fig. \ref{tau}(d)). We increase the laser peak amplitude to $a_L=20$ and the foil thickness to 200 nm. The plasma density is $n_e=100n_c$. As shown in Fig. \ref{tau}(e) and \ref{tau}(f), few-cycle XUV pulse emission with similar temporal and spectral signatures can still be found. The peak electric field of the XUV pulse is up to 8 TV$\mathrm{m}^{-1}$, which is about one order of magnitude smaller than the laser electric field of 60 TV$\mathrm{m}^{-1}$.

\section{Conclusion}

In conclusion, we have identified a new mechanism to generate intense single attosecond XUV pulse. The mechanism is associated with collective electron oscillation in plasmas. Making use of unipolarlike or subcycle laser pulses, a strong transverse transient net current can be excited in the plasma after the laser pulses, which is responsible for the XUV pulse emission. The emitted XUV pulse has both good temporal and spectral characteristics. Firstly, it is ultrashort in the time domain. The radiation is an isolated few-cycle attosecond XUV pulse with duration of several hundred attoseconds. Secondly, it has a narrow bandwidth in the spectral domain compared to other broadband XUV pulses. It has most energy confined around the plasma frequency and no low-harmonic orders below the plasma frequency. Thirdly, it has a high intensity. XUV pulse with peak electric field strength up to $ 8\times 10^{12} $ V$\mathrm{m}^{-1}$ has been generated. Moreover, it has a steep rise front. Such an intense attosecond few-cycle XUV pulse, without the need for pulse selecting and spectral filtering, is potentially useful for probing ultrafast attosecond dynamics and dense plasma properties, and also for applications of nonlinear optics in the XUV region.

\section*{Acknowledgments}

The first author would like to gratefully thank Dr. Hui-Chun Wu for valuable discussions. This work was supported in part by the National Natural Science Foundation of China
(Grant Nos. 61205100 and 11204281) and Key Foundation of CAEP (Grant No. 2012A0401016).

\end{document}